\documentclass{elsart5p}
\usepackage{graphicx}
\usepackage{amssymb}
\begin{document}

\begin{frontmatter}

\title{A sextupole ion beam guide to improve the efficiency and beam quality at IGISOL}

\author[1]{P. Karvonen},
\author[1]{~I.D. Moore\corauthref{cor1}},
\author[1]{~T. Sonoda\thanksref{thx1}},
\author[1]{~T. Kessler},
\author[1]{~H. Penttil\"{a}},
\author[1,2]{~K. Per\"{a}j\"{a}rvi},
\author[1]{~P. Ronkanen},
\author[1]{~J. \"{A}yst\"{o}}
\address[1]{Department of Physics, University of Jyv\"{a}skyl\"{a}, P.O. Box 35 (YFL), FI-40014 Jyv\"{a}skyl\"{a}, Finland}
\address[2]{STUK - Radiation and Nuclear Safety Authority, P.O. Box 14, FI-00881 Helsinki, Finland}
\corauth[cor1]{Corresponding author. Tel: +358 14 2602430; fax: +358 14 2602351.
               E-mail address: iain.moore@phys.jyu.fi (I.D. Moore).}
\thanks[thx1]{Present address: Atomic Physics Laboratory, RIKEN, 2-1 Hirosawa, Wako, Saitama 351-0198, Japan.}

\begin{abstract}
The laser ion source project at the IGISOL facility, Jyv\"{a}skyl\"{a}, has motivated the development and construction of an rf sextupole ion beam guide (SPIG) to replace 
the original skimmer electrode. The SPIG has been tested both off-line and on-line in proton-induced fission, light-ion and heavy-ion induced fusion-evaporation reactions 
and, in each case, has been directly compared to the skimmer system. For both fission and light-ion induced fusion, the SPIG has improved the mass-separated ion yields by 
a factor of typically 4 to 8. Correspondingly, the transmission efficiency of both systems has been studied in simulations with and without space charge effects. The 
transport capacity of the SPIG has been experimentally determined to be $\sim$10$^{12}$ ions s$^{-1}$ before space charge effects start to take effect. A direct comparison with the 
simulation has been made using data obtained via light-ion fusion evaporation. Both experiment and simulation show an encouraging agreement as a function of current 
extracted from the ion guide.
\end{abstract}

\end{frontmatter}

\section{Introduction}
\label{intro}
The ion guide technique developed at the University of Jyv\"{a}skyl\"{a} (JYFL) is an attractive method of producing exotic radioactive ion beams. It was originally 
conceived in order to overcome the long release times of the conventional ISOL approach, combined with the inability to produce refractory elements \cite{1}. At IGISOL, the projectile beam hits a thin target and the reaction product nuclei recoil out into a fast-flowing buffer gas, usually helium. As the ions slow down and thermalize their charge state changes continuously via charge exchange processes with the gas atoms and impurities within the gas. A significant fraction, typically 1-10\%, reaches a singly-charged state. This fraction along with all other species is transported out of the ion guide with the gas flow, whereby the ions are skimmed from the neutral gas and are injected into the mass separator via stages of differential pumping. After acceleration to between 30- and 40 kV depending on the experimental requirements the beam is mass separated by a dipole sector magnet, allowing separation of nuclei and contaminants with a typical mass resolving power of the order of 250. The attractive features of this technique are the fast ($\sim$milliseconds) evacuation times resulting in a chemical insensitivity of the ion guide, and the universality of the production method making it possible to produce even the most refractory of elements. 

In order to improve the elemental selectivity of the ion guide technique and in some cases the efficiency, a laser ion source project, FURIOS (Fast Universal Resonant laser IOn Source), is under development \cite{2}. Several laser ionization techniques are being developed. One is similar to the laser resonance ionization ion guide concept, IGLIS \cite{3,4}, where pulsed lasers are used to selectively ionize neutral atoms within the gas cell volume \cite{5}. In this method the lasers beams are transported into the ion guide either through a window on the rear of the cell or on the side, close to the exit hole. A second technique uses counter-propagating lasers beams to selectively ionize atoms after they have flowed out of the gas cell, within a radiofrequency guide located immediately after the ion guide \cite{6}. This is a variant of the so-called laser ion source trap (LIST) method designed to be installed at the Resonance Ionization Laser Ion Source facility, ISOLDE, CERN \cite{7}. By repelling the non-neutral fraction at the entrance to the radiofrequency guide, any ion transported to the extraction and acceleration stage of the mass separator is guaranteed to be a resonantly produced ion and therefore extremely high selectivity can be obtained.

When not being used in a LIST-type mode, the radiofrequency guide simply replaces the conventional skimmer system and is used in an ion transport mode. In this paper we concentrate on the use of the radiofrequency guide as a means to transport high current beams from the ion guide. Comparisons will be made to the skimmer-IGISOL system using both simulations and experimental data. Recently the effect of the baseline vacuum pressure in the IGISOL chamber on the evacuation time profiles of ionic yttrium and associate molecules has been studied using the laser ion source \cite{5}. Yttrium is an element that has a strong reactivity with water and oxygen impurities and very quickly reacts to form a chemical bond \cite{8}. In that work questions were raised with regards to the time of flight through the radiofrequency guide in light of the fast molecular formation reactions occurring either within the guide or its immediate surroundings when the baseline vacuum pressure was poor. This work additionally provides both simulation studies and experimental data in order to facilitate the understanding of the experimental results in \cite{5}.

\section{Radiofrequency ion beam guide}
\label{sec:1}

The idea of using a radiofrequency (hereafter called rf) multipole (sextupole) to transport radioactive ions extracted from an ion guide into a region of high vacuum was first developed and tested by Xu et al. \cite{9}. A similar system was adapted by Leuven to guide ions from the Ion Guide Laser Ion Source \cite{10}. The motivation for using a sextupole (SPIG) structure rather than a more commonly used quadrupole element stems from the capability of transporting higher current beams. The potential well created by the rf potential associated with a SPIG does not have such a sharp minimum as that of a quadrupole and the potential walls are higher and steeper. The maximum depth of the pseudopotential formed by the rf multipole field in a guide with the same geometry and rf voltage is proportional to N$^{2}$, where there are 2N parallel rods (N = 3 in the case of a SPIG). 

The use of a sextupole ion beam guide at the IGISOL facility was previously reported in \cite{11}. It was compared to a skimmer-ring system in experiments involving both fission and heavy-ion fusion-evaporation reactions. The geometry of the SPIG was smaller than that to be discussed in the present work, and the motivation for its use was rather different, primarily to reduce the beam energy spread of the skimmer-ring system (100 eV) to the level of approximately 1 eV. In that work, the production yield of the HIGISOL (Heavy-ion Ion Guide Isotope Separator On-Line) reaction, $^{94}$Mo($^{36}$Ar,3p2n)$^{125}$La, was only slightly improved with the use of the SPIG (a factor of 1.3$\times$), whereas in the fission reaction,$^{238}$U(p,f)$^{112}$Rh, the yield of $^{112}$Rh improved by a factor of five. One of the most important improvements however, was the resultant increase in the mass resolving power by a factor of typically five thus reducing the mass contamination in the focal plane of the mass separator and therefore increasing the sensitivity of the study of the most exotic nuclei. It was concluded that the designed SPIG could transport beam intensities of the order of 10$^{12}$ ions s$^{-1}$ before becoming unstable.

The development of the laser ion source in Jyv\"{a}skyl\"{a} has motivated the design and construction of a new SPIG to be used for the JYFL LIST project. The high-repetition rate (10 kHz) laser system combined with the gas velocity immediately after the exit hole of the ion guide ($\sim$1000 ms$^{-1}$) results in a required interaction length between the neutral atoms and laser light of  $>$10 cm if each atom is to have a chance of interacting with the laser beams at least once. The typical size of the laser beam diameter is $\sim$6 mm, tapering to a focus of 2 mm at the exit hole of the ion guide. Both the length and inner diameter of the SPIG therefore have some constraints due to the spatial and temporal properties of the pulsed lasers. A further technical issue to address is the need for an efficient overlap of the laser beam and the gas jet exiting the ion guide. Off-line studies of the shape of the gas jet have been extensively performed at Jyv\"{a}skyl\"{a} using  both visual observations and the transport of isotopes recoiling from a $^{223}$Ra source. One result from these studies is that the background pressure in the region of the expanding jet is an important parameter \cite{12,13}. An increase in the background pressure leads to a narrower gas jet, and hence an improved overlap with the counter-propagating laser beams.

\begin{figure}[ht]
\begin{center}
\includegraphics*[width=9cm]{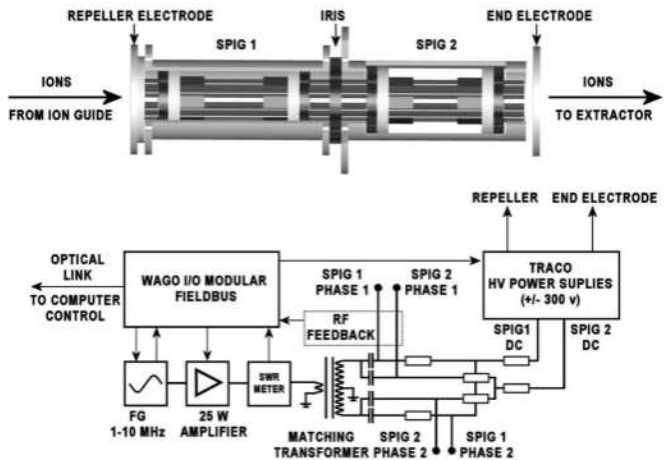}
\end{center}
\caption{Schematic design of the sextupole ion beam guide (SPIG) for use in the LIST project. FG = function generator.}
\label{fig:1}
\end{figure}

A schematic representation of the rf sextupole ion beam guide designed for the LIST project is shown in Fig. 1 and a complete list of dimensions and operational
parameters is given in Table 1. The SPIG has been constructed in two segments, seen most clearly in the photograph of Fig. 2. The first has an enclosed structure around the parallel rods in order to keep the background pressure high enough to collimate the gas jet. The second SPIG is ``open'' and any remaining buffer gas or neutral fraction can be efficiently pumped away through the gaps between the rods. An adjustable stainless steel iris separates the two segments, added to control the pressure in a coarse way in the first segment. It has not, however, been used in this work and therefore is kept fully open (22 mm inner diameter). 

A function generator produces a sinusoidal waveform of variable frequency which is amplified by a 25 W amplifier and impedance matched to the SPIG rods. Two transformers are available with different coil winding ratios when the SPIG needs to be operated with slightly different rf frequencies, however in general the SPIG is rather insensitive to the choice of coil and 3 MHz is typically used. DC potentials can be added to the rf signal to generate small accelerating potentials within the SPIG. The impedance matching circuit and transformer coils sit at high voltage in the high radiation IGISOL chamber area and therefore full automation of the SPIG device is necessary. The SPIG parameters (dc and rf) can be adjusted and monitored by a labview-based control program. One of the most important features of the program is that it monitors the reflected and transmitted rf power to the SPIG. During initial tuning of the SPIG it is sometimes necessary to optimize the rf frequency to ensure the reflected power is minimal, hence the impedance is well-matched. Related to this, the temperature of the rf amplifier is also monitored.

\begin{table}[h]
\caption{Dimensions and operational parameters of the JYFL SPIG.}
\label{tab:1}
\begin{tabular}{ll}
\hline\noalign{\smallskip}
Dimension & Value [mm] \\
\noalign{\smallskip}\hline\noalign{\smallskip}
Repeller aperture & 6 \\
Repeller thickness & 3 \\
SPIG inner diameter & 10 \\
SPIG rod diameter & 4 \\
Iris aperture & 0 - 22 \\
SPIG 1 axial length & 78.5 \\
SPIG 2 axial length & 81.5 \\
Total axial length of the device & 165.5 \\
End electrode aperture & 6\\
\noalign{\smallskip}\hline
\hline\noalign{\smallskip}
Parameter & Value \\
\noalign{\smallskip}\hline\noalign{\smallskip}
RF frequency & 3 or 4 MHz \\
RF amplitude & 0 - 600 V$_{pp}$ \\
Repeller voltage & 0 $\pm$300 V \\
SPIG rod dc voltages & 0 - 300 V (-V$_{e}$) \\
End electrode voltage & 0 - 300 V (-V$_{e}$)\\
\noalign{\smallskip}\hline
\end{tabular}
\vspace*{2pt}
\end{table}

A separate electrode can be added to the front end of the SPIG if needed. This first electrode, the repeller electrode, can be set to either positive or negative potential (0 - $\pm$300 V) with respect to the ion guide. The typical operating distance between the exit hole of the guide and the repeller electrode is 5 mm. When the SPIG is used in the LIST mode then the electrode can be biased to repel any non-neutral fraction exiting the gas cell. A final ``end electrode'' located after the second SPIG segment is used to optimize the ion transport through the SPIG however it can also be set to a potential that is positive with respect to the SPIG dc levels. This results in a confinement of the ions and a means to bunch if required, though with the installation of the gas-filled radiofrequency quadrupole cooler/buncher device downstream from the mass separator this operation has not been necessary [14]. Collisional cooling between the ions and the gas atoms can happen within the confinement of the SPIG if the kinetic energy of the ions is higher than the thermal energy of the gas. The ions are finally transported to the extraction electrode (not shown in Fig. 1) at a distance of 20 mm behind the final SPIG electrode. The acceleration of the ions at the extraction electrode occurs in a region far from the high pressure zone in which the skimmer electrode previously operated and consequently the energy spread and spatial extent of the ions is small.
	
\begin{figure}[h]
\begin{center}
\includegraphics*[width=9cm]{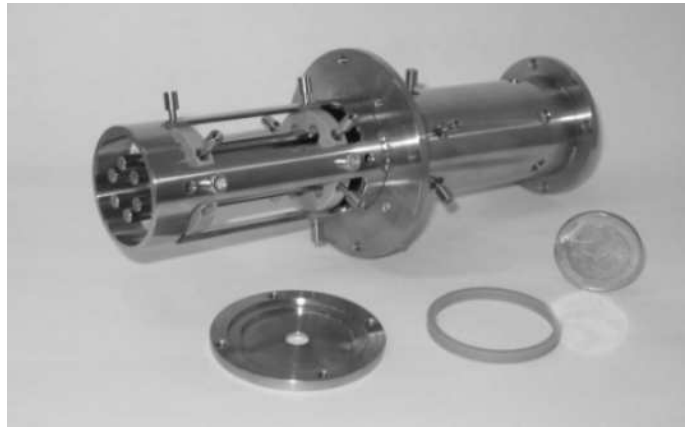}
\end{center}
\caption{Photograph of the SPIG (colour on-line) shown relative to the size of two euro coin. The first segment is enclosed to keep the environmental pressure higher for a narrowing of the gas jet. The second segment is open.}
\label{fig:2}
\end{figure}

\section{Monte Carlo simulations}
\label{sec:2}
Monte Carlo simulations have been performed in order to gain a better understanding of the properties of the SPIG and to extract important parameters such as the emittance of the ion beam and efficiency of transmission, in particular in light of the promising improvements compared to the skimmer-ring system discussed in \cite{11}. The simulations described in this work have been realized using the SIMION 3D simulation software package [15] and are discussed in the following separate sub-sections.

\subsection{Description of simulation code}
The Runge-Kutta method has been utilized to model ion trajectories under the influence of an electric field, and the effect of the helium gas flow has been treated using a modified version of the hard-sphere collision model [16], in which the individual collisions between the ions and gas atoms (hard spheres) are modeled with random collision angles. The number of collisions and the corresponding mean free path of the ions are determined by the input parameters which include pressure, temperature and a collision cross section of the colliding particles. The background gas is assumed to have a non-zero mean velocity, with a Maxwell-Boltzmann distribution of velocities as a function of temperature. One important missing piece of information is a direct measurement of the background gas pressure within the SPIG. The pressure and velocity vectors have therefore been based on earlier simulations using the CosmosFloWorks package \cite{17}.

\begin{figure}[h]
\begin{center}
\includegraphics*[width=9cm]{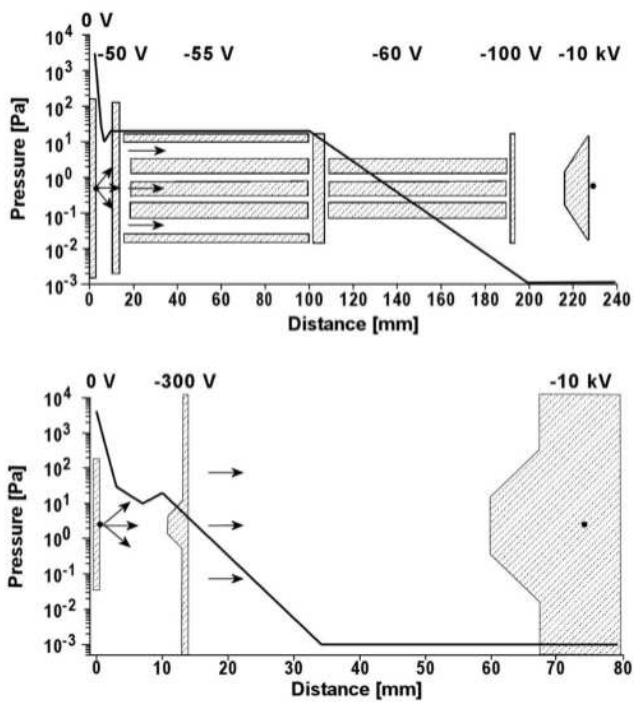}
\end{center}
\caption{Simulation schematics for the SPIG (top) and for the skimmer (bottom). The solid lines show the simulated trend of the buffer gas pressure while the arrows 
indicate the velocity vector directions of the buffer gas model. The ions are created in a circular source (1.2 mm diameter) in front of the ion guide on the left of both figures. The point of detection is indicated by a spot after the extractor electrode. The ion guide pressure is set to 100 mbar (10$^{4}$ Pa). Typical electrode dc potentials are indicated. The rf frequency is 3 MHz and rf amplitude 250 $V_{pp}$ unless otherwise stated.}
\label{fig:3}
\end{figure}

The collision cross section is simply the sum of the diameters of the colliding particles squared. In this work the Van der Waals diameter of the particle has been used. This is calculated from experimental data of the atomic spacing between pairs of unbound atoms in a crystal and it can be understood as representing the diameter of an imaginary hard sphere. Two atoms of different masses have been chosen to act as ``test'' ions in the following simulations, $^{20}$Ne and $^{106}$Pd, with diameters of 1.54 \AA \ and 1.63 \AA \ respectively \cite{18}. The simulation time step is determined primarily by the hard-sphere collision model and is related to the background gas pressure. The only exception is when the time step is larger than that needed for a reliable generation of the rf field within the model. This may happen if the background pressure is very low, and hence the mean free path between collisions is large. In this case the time step is forced to take the value of the rf cycle time divided by 200. 

A source of 500 ions has been randomly generated within a circular disk of equivalent diameter to the ion guide exit hole. The initial buffer gas velocity vectors are modeled in an ``explosion-like'' scenario restricted to a 60$^{o}$ solid angle between the ion guide and the first element of the simulated devices in order to mimic the forward focusing of the gas jet. After the first element only the optical axis direction is included in the simulations (both gas pressure and flow velocity are then one dimensional), the single ion trajectories are followed and are finally recorded after passing through the extractor electrode (diameter 7 mm). A schematic of the simulation for both the SPIG and skimmer is shown in Fig. 3. Simulation geometries and other parameters such as dc and rf amplitude have been selected to correspond to typical experimental values. In the following simulations (sections 3.1 to 3.4) effects related to space charge and electrodes downstream from the extractor electrode have been neglected. Space charge will however be treated separately in section 3.5. The modeled behavior of the buffer gas pressure and velocity is rather similar in both cases (Fig. 3) and exponentially decreases as a function of the distance from the ion guide, except in the region inside the first SPIG element where it remains constant due to the enclosure. The rather erratic solid line representing the buffer gas pressure between the ion guide and skimmer electrode (less so between the guide and SPIG repeller electrode) represents areas of turbulence and normal shocks seen in the earlier CosmosFloWorks simulations and previous visual observations.

\subsection{Beam emittance}
Figure 4 shows the resulting root mean square emittance obtained as a function of ion guide pressure for \textit{A} = 20 and \textit{A} = 106. The quality of the ion beam after the extractor is clearly sensitive to the gas pressure for both the skimmer and SPIG devices however the trends they follow as a function of ion guide pressure are very different. The skimmer produces ion beams with the lowest emittance (highest quality) when the buffer gas pressure is low. The typical on-line operating conditions at IGISOL correspond to ion guide pressures close to 300 mbar, and thus although the emittance improves at lower pressures the stopping efficiency of recoil products decreases. An increase in the buffer gas pressure results in a higher number of collisions between the ions and gas atoms during acceleration by the electric field. This in turn leads to a poorer quality ion beam, in particular for lighter mass ion beams. The volume between the skimmer and extraction electrode where the ions are accelerated rapidly is most critical for collisional scattering. By increasing the baseline pressure in this region (see the trend in Fig. 3) the emittance increases and, at some point, the ion beam will start to be lost due to collimation by the extractor electrode.

\begin{figure}[h]
\begin{center}
\includegraphics*[width=9cm]{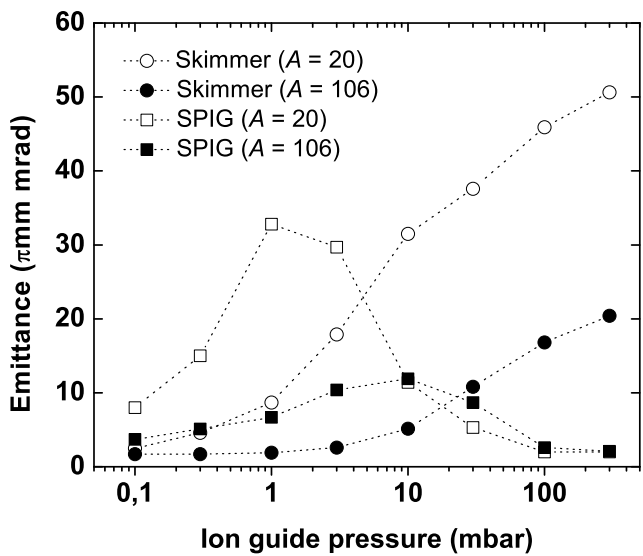}
\end{center}
\caption{A comparison between the simulated ion beam emittances for ions of mass \textit{A} = 20 and \textit{A} = 106 as a function of ion guide pressure for the SPIG and skimmer devices.}
\label{fig:4}
\end{figure}

The SPIG on the other hand exhibits the feature of buffer gas cooling (discussed later in connection with Fig. 7) with increasing ion guide pressure, translating directly into a reduced beam emittance for beams of masses heavier than the buffer gas (in this case helium). This is most clear for ion beams of lighter masses \textit{A} = 20 where cooling occurs at lower gas pressures than for heavier masses. The effect of improved beam quality when replacing a skimmer with a SPIG has been documented previously \cite{10,11} via measurements of the mass resolving power (MRP, defined as M/$\Delta$M where M is the separated mass and $\Delta$M  the width of the mass peak at FWHM) of the mass separator in the focal plane. In the Leuven Isotope Separator On-Line, LISOL, resonantly laser ionized nickel atoms from a heated filament were used to measure a MRP of order 300 when the skimmer was installed and values as high as 1450 were obtained with a SPIG \cite{10}. In earlier work at IGISOL the MRP of a SPIG was measured to be at best 1100 with a $^{223}$Ra $\alpha$-decay recoil source, obtained with a rather small voltage of -20 V between the end plate and the sextupole rods \cite{11}. It was noted that the scattering of ions being accelerated between the SPIG rods and the end plate contributed importantly to beam quality, due to the relatively high pressure in that region. This MRP can be compared to a typical IGISOL-skimmer value of 250 \cite{19}.

\subsection{Transmission efficiency}

	The effect of the ion guide pressure on the transmission efficiency of the skimmer and SPIG systems is illustrated in Fig. 5. The efficiency with the skimmer drops
rapidly for lighter masses as the ion guide pressure increases, whereas for heavier masses (\textit{A} = 106) the efficiency is constant and then starts to decrease after $\sim$10 mbar. The losses are related to the increasing number of collisions between the ions and the buffer gas atoms reducing the number of ions that pass through either the skimmer hole or the extractor electrode aperture, due to a fixed geometrical acceptance. According to the model, the shoulder visible between 10 and 30 mbar gas pressure in the A = 20 skimmer efficiency curve is not due to statistical fluctuations. An increasing number of buffer gas collisions in the region around the skimmer limit the radial motion of the ions in such a way that they can be more efficiently guided through the electrode by the electric field. This leads to a saturation of the transmission efficiency of the electrode above 30 mbar ion guide pressure. As the pressure increases however, the background pressure between the skimmer and extractor becomes worse and thus the transmission efficiency through the extractor reduces, resulting in a continuing decrease in the total transmission efficiency at higher gas pressures.

\begin{figure}[h]
\begin{center}
\includegraphics*[width=9cm]{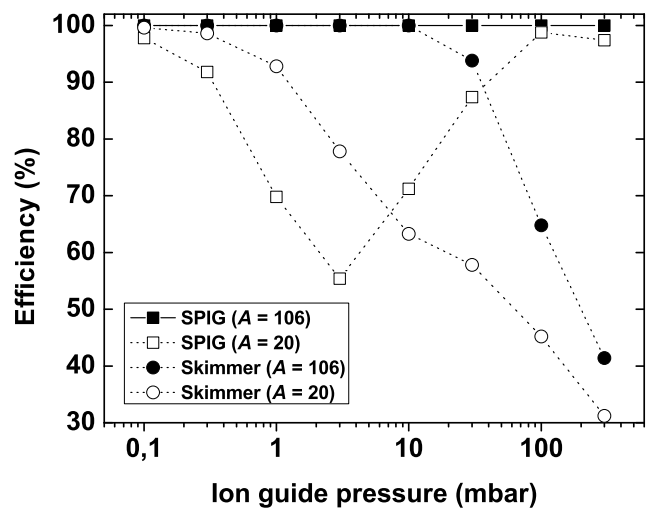}
\end{center}
\caption{A comparison between the simulated transmission efficiencies for ions of mass \textit{A} = 20 and \textit{A} = 106 as a function of ion guide pressure for the SPIG and skimmer devices.}
\label{fig:5}
\end{figure}

The ratio of the loss fraction due to collisions on the skimmer electrode compared to the extractor has been studied for \textit{A} = 20 as a function of ion guide pressure, with the total efficiency corresponding to the trend shown in Fig. 5. At 30 mbar pressure and -300 V on the skimmer, 95\% of the total losses occur on the skimmer electrode with 5\% on the extractor. This ratio gradually reduces as the pressure increases. At 300 mbar, 73\% of the total losses occur on the skimmer and 27\% on the extractor electrode. The loss on the skimmer electrode can be recovered by increasing the skimmer voltage, however at the cost of increasing losses on the extractor. For example, by increasing the skimmer voltage to -400 V at 300 mbar ion guide pressure, 67\% of the total ions lost occur on the skimmer with 33\% lost on the extractor. Figure 6 illustrates the ion trajectories for the skimmer system with -300 V applied and 300 mbar for both \textit{A} = 20 and \textit{A} = 106. In the latter case it can be seen that all losses occur on the skimmer electrode.

The initial offset of the SPIG transmission efficiency for ions of mass \textit{A} = 20 at 0.1 mbar ion guide pressure compared to the skimmer is due to the reduced SPIG performance for low mass ion beams without the effect of buffer gas cooling. It has been seen in the simulation that at low ion guide pressures, rare collisions can cause unstable trajectories for light ions, particularly if the SPIG operates close to the instability point. This is the reason why the transmission efficiency drops as the pressure increases from 0.1 mbar to 3 mbar. As the pressure continues to increase buffer gas cooling takes effect and the efficiency increases to a level of $>$90\%. At 300 mbar pressure, the efficiency again drops for light masses, due to losses both within the SPIG and on the extraction electrode. This can be compared to \textit{A} = 106 in which the transmission efficiency is constant at 100\% for the SPIG system.

\begin{figure}[h]
\begin{center}
\includegraphics*[width=9cm]{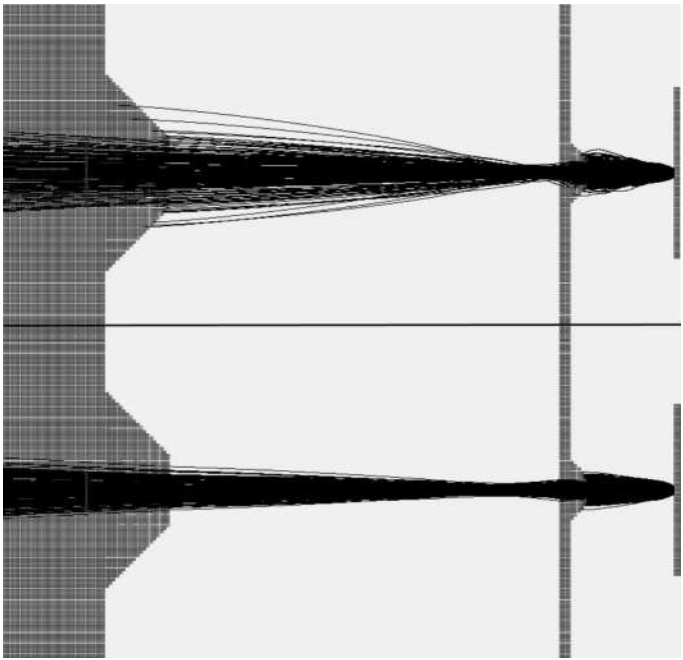}
\end{center}
\caption{Ion trajectories of \textit{A} = 20 (top) and \textit{A} = 106 (bottom) through the skimmer and extractor electrode at a skimmer voltage of -300 V and an ion guide pressure of 300 mbar. The ion guide is on the right of the figure.}
\label{fig:6}
\end{figure}

\subsection{Total kinetic energy}

	Figure 7 shows the total kinetic energy of the ions in the SPIG system as a function of the z-axis distance from the ion guide. In this simulation the ion guide 
pressure was fixed at 200 mbar and the rf amplitude 200 V$_{pp}$. The positions of the relevant electrodes are shown with respect to the ion guide. From the creation 
point, the ions are accelerated by the potential difference between the ion guide (assuming gas flow alone, no internal field guidance within the ion guide) and the 
repeller electrode. As the ions enter the first enclosed SPIG structure they start to cool down via collisions with the gas atoms. By the time the ions reach the position of the iris they have thermalised. Here the ions are again accelerated due to the dc voltage of SPIG 2. Within the second SPIG they effectively maintain a constant level of kinetic energy and drift towards the end electrode. At a distance of approximately 180 mm from the ion guide the ions pass the end electrode and are accelerated to the full potential of the extractor electrode (-10 kV).

\begin{figure}[h]
\begin{center}
\includegraphics*[width=9cm]{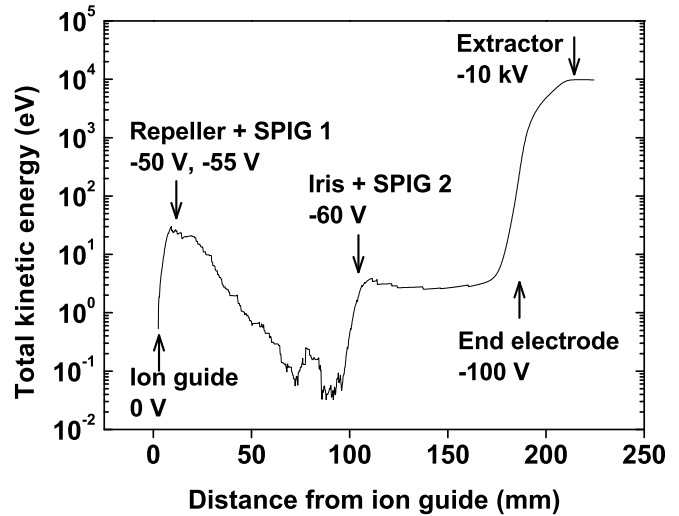}
\end{center}
\caption{Simulated total kinetic energy of an ion with mass \textit{A} = 106 as a function of optical axis distance from the ion guide. The electrode potentials and relative positions with respect to the ion guide are labeled in the figure.}
\label{fig:7}
\end{figure}

In the region of the extractor nozzle we can assume there are no more collisions between the buffer gas atoms and the ions of interest. The total kinetic energy shown in Fig. 7 also has an associated standard deviation. This value, $\Delta$KE, multiplied by the root mean square value of the x-axis position (radial axis) is proportional to the mass resolving power. For the data shown in Fig. 7, $\Delta$KE = 20 eV and X$_{rms}$ = 0.39 mm. The corresponding data for the skimmer system is $\Delta$KE = 89 eV and X$_{rms}$ = 0.82 mm. Therefore, according to the simulation, the mass resolving power of the SPIG is expected to be $\sim9 \times$ higher than that of the skimmer.

\subsection{Time of flight}

The time of flight of ions with mass \textit{A} = 20 and \textit{A} = 106 through the SPIG was extracted as a function of ion guide pressure. Figure 8 shows the results obtained from the simulation firstly using the baseline pressure trend within the SPIG as shown in Fig. 3 and, secondly, assuming one order of magnitude improvement in the baseline pressure throughout the SPIG system while maintaining the same initial ion guide pressure. The results illustrate the sensitivity to the background pressure environment on the transport time of ions of different mass through the SPIG. If the buffer gas pressure is set to 0 mbar then the time of flight through the SPIG is determined solely by the dc and rf parameters. This condition can be assumed for ions that are extracted from the SPIG on-axis. For \textit{A} = 20 the flight time is estimated to be 8.5 $\mu$s and for \textit{A} = 106, a value of 19.3 $\mu$s is extracted. These values are equivalent to those illustrated for the two masses at the lowest simulated ion guide pressures of Fig. 8 and simply reflect the insensitivity of the time of flight of ions to the background pressure at such low ion guide pressures (and therefore such low baseline pressures).

\begin{figure}[h]
\begin{center}
\includegraphics*[width=9cm]{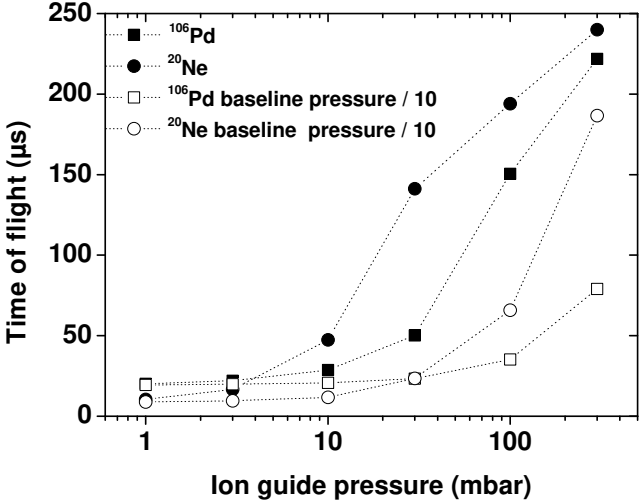}
\end{center}
\caption{Simulated time of flight (tof) of ions with mass \textit{A} = 20 and \textit{A} = 106 through the SPIG as a function of the ion guide pressure for two different baseline pressure conditions. The data points represent the average value of 361 generated ions.}
\label{fig:8}
\end{figure}

\subsection{Effects of space charge}
The simulations discussed in sections 3.2 and 3.3 are important to understand the intrinsic differences of the skimmer and SPIG systems, considering the geometrical 
acceptance of such devices are very different. Here we introduce in a simple model the effect of transporting higher current beams from the ion guide through the devices in order to understand better the effect on the transmission efficiency. Experimentally, currents measured after the devices typically vary from several tens to several hundreds of nA. As the current increases it can disturb the applied electric fields causing space charge effects that can limit the transmission efficiency. The principle behind the simulation is to calculate the repulsive electric field produced by a pre-defined amount of charge generated by the output current from the ion guide. The repulsive field is then coupled to the simulation together with the buffer gas model described in section 3.1. Trajectories of 500 test ions are followed and the transmission efficiency is determined as a function of the background ion current.  
Accurate modeling of the charge density, in particular in the case of the SPIG, is rather difficult because parameters such as charge radius and density are coupled not only to the environmental parameters but also to the total amount of charge itself. The realistic total amount of charge is not uniformly distributed over a simulated volume. Instead, the charge density varies as a function of ion velocity providing regions of high charge density in places where the ion velocity is low, and less dense regions where a high acceleration potential is applied. In mathematical terms this means that there is no symmetrical Gauss surface available for modeling the effect of charge.

Some assumptions have been made in order to simplify the model and to make it easier to adapt to the simulation software. Firstly, it is assumed that the test ion can only feel the effect of the external charge when it is located within a very short axial range. This means that the charge density can be kept constant in that range.
Secondly, it has been assumed that the charge inside the range is uniformly distributed over a cylindrical volume centered axially on the test ion. With these two assumptions the axial components of the electric field generated by the external charge are cancelled and cylindrical symmetry can be used over a short range. The repulsive force is calculated by modeling the external charge from the ion guide as a series of infinitely-long homogeneously-charged cylinders. Each cylinder has a
position-dependent radius R(z) and a ``charge per unit length''- factor which is matched to correspond to the charge density calculated from  a fixed total ion beam current divided by the local axial velocity of the test ion (Eq. (1)). In the SPIG, test ions may be temporarily stopped because of cooling and buffer gas collisions leading to unrealistically high repulsive electric fields. This is overcome by defining a minimum velocity that can be used corresponding to the kinetic energy of the cooled ion inside the first SPIG element. The repulsive electric fields are then modeled using
 \begin{equation}
  E(r,z)=\frac{Ir}{2\pi\epsilon_{0}\nu_{z}(z)R(z)^{2}} \ \ r<R
 \end{equation}
 \[
	E(r,z)=\frac{I}{2\pi\epsilon_{0}\nu_{z}(z)r} \ \ r\geq R
 \]	
where \textit{r} is the ion radius from the optical axis, \textit{I} is the total current from the ion guide and \textit{v$_{z}(z)$} is the ion axial velocity at the centre of the cylinder.

When simulating the ion beam transmission through the skimmer, the position-dependent radius of the charged cylinder follows the geometrical acceptance of the elements. This means that the background charge forms a cone of length 10 mm (the distance from the ion guide to the skimmer nozzle) with an initial radius of 0.6 mm (matching that of the exit hole) increasing to a radius of 1 mm which matches the skimmer nozzle. The path from the skimmer to the extractor is generated in a similar manner. In the case of the SPIG, the cone is formed only in the region between the ion guide and the repeller electrode, with a final radius of 3 mm. After the repeller electrode a constant radius of 4.5 mm is used throughout the rest of the system (recall the inner diameter of the SPIG is 10 mm). For a given ion guide output current, the current within the SPIG is determined once the losses between the ion guide and repeller have been taken into account. An ion guide pressure of only 100 mbar was used in the simulations in order to provide faster simulation times. To improve the overall efficiency of both devices, the axial acceleration has been increased. A skimmer voltage of -500 V was applied and, for the SPIG simulation, the dc potential of all electrodes was lowered by 50 V. The SPIG rf amplitude was also increased to 250 V$_{pp}$ which is closer to the typical experimental value. 

An example of the ion trajectories for \textit{A} = 106 through the SPIG system with an ion guide output current of 1 $\mu$A is shown in Fig. 9. The initial cone of charge created at the exit hole of the ion guide fills the aperture of the repeller electrode. This is the region where the majority of ion losses occur. Similarly, Fig. 10 illustrates the effect on the ion trajectories in the skimmer system when the ion guide output current is increased to 4 $\mu$A. This highlights the extreme case in which only one ion is successfully transported to the extractor electrode. All other ions are lost on the skimmer electrode due to the intense space charge effect.

\begin{figure}[h]
\begin{center}
\includegraphics*[width=9cm]{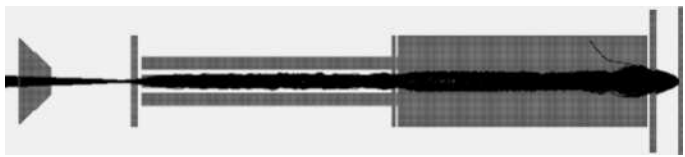}
\end{center}
\caption{Ion trajectories of \textit{A} = 106 through the SPIG and extractor electrode with an ion guide pressure of 100 mbar, a SPIG rf amplitude of 250 V$_{pp}$ and an ion guide output current of 1 $\mu$A. The ion guide is on the right of the figure.}
\label{fig:9}
\end{figure}

\begin{figure}[ht]
\begin{center}
\includegraphics*[width=9cm]{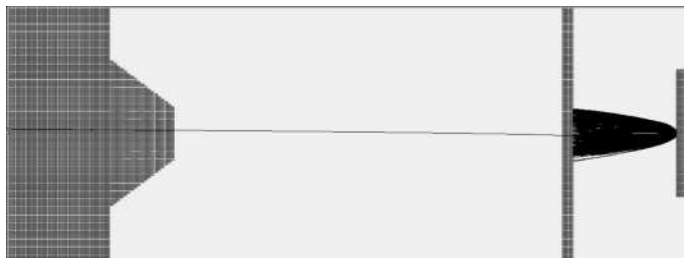}
\end{center}
\caption{Ion trajectories of \textit{A} = 106 in the skimmer system with an ion guide pressure of 100 mbar, a skimmer potential of -500 V and an ion guide output current of 4 $\mu$A. Only one ion in 500 test ions is successfully transmitted through the extractor.}
\label{fig:10}
\end{figure}

The transmission efficiency for \textit{A} = 20 and \textit{A} = 106 ions as a function of ion guide current is shown in Fig. 11. At low currents ($\sim$10 nA) both devices are in saturation, no space charge exists and in general the efficiencies compare rather well with those of Fig. 5 taken at 100 mbar ion guide pressure. The deviation of the skimmer efficiency for \textit{A} = 106 ions in Fig. 11 ($\sim$80\% efficiency) with that of Fig. 5 ($\sim$65\% efficiency) is due to the increased skimmer voltage in the space charge simulations. This increase in potential has little effect on the \textit{A} = 20 ions as they are extracted with optimal efficiency at a lower skimmer voltage. The results for a low ion beam current suggest that the geometrical acceptance dominates the transmission efficiency, whether due to losses on the skimmer or SPIG, or to losses on the extraction electrode. As the beam current increases both devices start to exhibit a drop in efficiency, albeit at different currents, as the guiding fields become distorted and the ions start to repel each other transversely. At a current of 1 $\mu$A the transmission efficiency of the skimmer has dropped to 20-30\% for \textit{A} = 106 and \textit{A} = 20, respectively. For the same beam current, the SPIG transmission efficiency has reduced to 40-55\% for \textit{A} = 106 and \textit{A} = 20, respectively.

\begin{figure}[h]
\begin{center}
\includegraphics*[width=9cm]{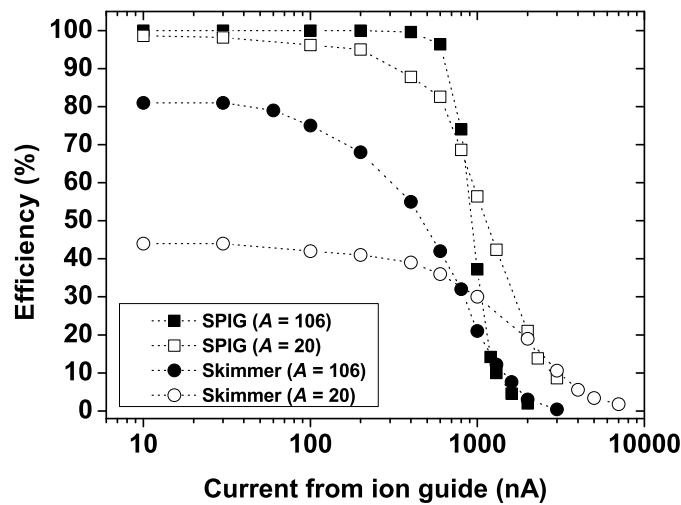}
\end{center}
\caption{Skimmer and SPIG transmission efficiencies as a function of the ion guide current output.}
\label{fig:11}
\end{figure}

\section{Experimental results}

This section consists of an accumulation of results from several experiments during a three year period from 2005 to 2008. Sections 4.1 to 4.4 deal with yield comparisons between the skimmer and SPIG devices both off-line and on-line, using a variety of reactions. Section 4.5 details recent efforts to obtain a value for the time of flight through the SPIG system motivated by recent studies of fast molecular formation reactions identified using the laser ion source. Finally section 4.6 summarizes experimental data obtained to extract the transmission efficiency of the SPIG as a function of ion guide current. A full description of the ion guides and associated reaction kinematics used in this work may be found elsewhere \cite{1,11}.

\subsection{$^{223}$Ra $\alpha$-decay recoil source}
The first measurements of the SPIG described in section 2 were performed in June 2005. An $\alpha$ decay recoil source of $^{223}$Ra \cite{11} was collected onto the tip of an aluminium rod and mounted within the volume of a discharge-type ion guide. Alpha recoils of its decay product, $^{219}$Rn, were ejected from the source and guided to the exit hole in a helium gas flow. The ions were transported either through the rf-sextupole or skimmer, mass separated and implanted into a foil in front of a silicon detector in the focal plane of the magnet. The efficiency was measured as a function of the helium gas pressure and includes a 30\% detector efficiency. The results can be seen in Fig. 12. The skimmer was operated at approximately -120 V with respect to the ion guide and was positioned at a distance of 10 mm from the exit hole. The SPIG rf amplitude was operated at 320 V$_{pp}$, with -40 V dc on SPIG 1 and -60 V dc on SPIG 2, with a 3 mm gap between the exit hole and the repeller electrode.

\begin{figure}[h]
\begin{center}
\includegraphics*[width=9cm]{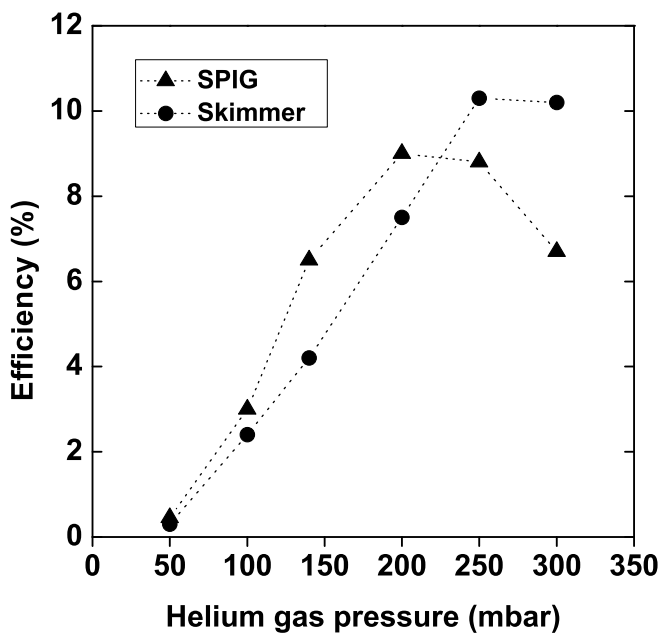}
\end{center}
\caption{The measured efficiencies of $^{219}$Rn$^{+}$ as a function of ion guide pressure using both the skimmer- and SPIG-IGISOL systems.}
\label{fig:12}
\end{figure}

At an ion guide pressure of 50 mbar the recoil range of $^{219}$Rn is $\sim$10 mm and therefore the ions can be lost due to collisions with the walls. It can be seen that the transmission efficiencies with the SPIG and skimmer are rather comparable. The SPIG may be consistently performing slightly better at lower ion guide pressures however there are no striking differences. The final point in the SPIG data at 300 mbar drops, yet as there are no data points at higher pressures no conclusions can be made. Similarly it is seen that the skimmer data appear to saturate, yet again no firm conclusions can be drawn as to whether this is related to the stopping of recoils inside the ion guide.

\subsection{Light ion fusion-evaporation reactions}

The light-ion fusion-evaporation reaction $^{40}$Ca(p,n)$^{40}$Sc was used to produce $^{40}$Sc ($T_{1/2}$ = 183 ms) at a proton bombarding energy of 35 MeV from the JYFL K-130 cyclotron. The target thickness was of the order of a few mg/cm$^{2}$. A silicon detector with an efficiency of 30\%, mounted behind a set of slits just after the focal plane of the mass separator, was used to count the positrons emitted in the decay of the $^{40}$Sc$^{+}$ reaction products. The skimmer and SPIG devices were both installed during the same run after some hours of ``cooling time'' between measurements. The ion yield as a function of the primary beam intensity for both the skimmer and SPIG system are shown in Fig. 13. In both cases the helium buffer gas pressure was fixed at an optimum value of 150 mbar for all primary beam intensity measurements. The voltage on the skimmer plate was optimized at -300 V with respect to the ion guide. The linear fits to the data in Fig. 13 shows that the SPIG provides on average 8 times higher yield of $^{40}$Sc as compared to the skimmer with same primary beam current.
The mass resolving power was measured at mass \textit{A} = 40 (argon) on a Faraday cup in the focal plane before the slits, rather than on the silicon detector, and yielded results of $\sim$250 and $\sim$620 for the skimmer and SPIG, respectively.

\begin{figure}[h]
\begin{center}
\includegraphics*[width=9cm]{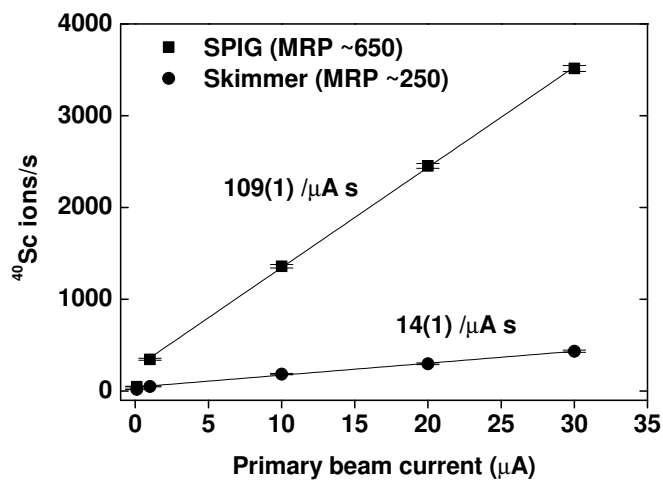}
\end{center}
\caption{$^{40}$Sc yields produced by the $^{40}$Ca(p,n)$^{40}$Sc reaction with a proton beam energy of 35 MeV. The solid lines represent the yield per $\mu$A of primary beam current estimated from a linear fit to the data points. Statistical errors are also shown combined with a detector efficiency of 30\%.}
\label{fig:13}
\end{figure}

During the same experiment the ion guide had a 3 mg/cm$^{2}$ $^{58}$Ni target enriched to a level of 99.8\% installed on the opposite side to the $^{40}$Ca target. This afforded a comparison of the reaction $^{58}$Ni(p,n)$^{58}$Cu between the SPIG and earlier experiments using the skimmer. The reaction product $^{58}$Cu ($T_{1/2}$ = 2.3 s) was produced at a bombarding energy of 18 MeV. The first reported yield measurement of $^{58}$Cu at IGISOL was performed using a ring-electrode-skimmer system, at what is now termed IGISOL 2 \cite{20}. At a primary beam intensity of 15 $\mu$A the yield in that work was 4000 ions s$^{-1}$. Following the upgrade of the IGISOL pumping capacity in early 2004 a new yield measurement was reported at ``IGISOL 3'' \cite{21}. The measurement was made at a primary beam intensity of 1 $\mu$A and the yield of $^{58}$Cu was given as 1500 ions/$\mu$A - an increase of 5.6 over the yield reported in \cite{20} assuming that the $^{58}$Cu yield increases linearly with primary beam intensity. The main improvement to the light-ion induced yields reported in \cite{21} appears to have been due to a better transmission through the extraction electrode which was increased in diameter from 4 mm to 7 mm, a factor of three in area. Indeed the simulations reported in this paper illustrate losses at higher ion guide pressures due to collisions between the ions and gas atoms in the region between the extractor and skimmer, suggesting a further increase in extractor diameter may provide additional improvements with the skimmer system. In the present work a measurement of the yield of $^{58}$Cu at 15 $\mu$A primary beam intensity afforded a direct comparison to that reported by Per\"{a}j\"{a}rvi \textit{et al}. \cite{20}. With the SPIG system, a yield of $\sim$48000 ions s$^{-1}$ was extracted from the data which is a factor of 12× improvement over the original ``IGISOL 2'' measurement (hence an improvement factor of 2.1 compared to the ``IGISOL 3'' yield extrapolated to 15 $\mu$A). A direct measurement of $^{58}$Cu at 1 $\mu$A primary beam intensity resulted in a yield of $\sim$6850 ions s$^{-1}$, a factor of 4.6× improvement over the value reported in \cite{21}. The discrepancy between the two improvement factors of the present work with respect to ``IGISOL 3'' may be due to an erroneous measurement of the yield of $^{58}$Cu reported in \cite{21}.

\subsection{Proton-induced fission reaction}

The fission ion guide is the most commonly used ion guide at IGISOL. One of the first on-line experiments using the SPIG involved mapping the independent fission yield distribution using JYFLTRAP \cite{22}. In this experiment a 30 MeV proton beam impinged on a $^{238}$U target tilted by 7 degrees with respect to the beam direction to increase the effective thickness from 15 mg/cm$^{2}$ to 120 mg/cm$^{2}$. A typical reaction product produced close to the peak of the fission mass distribution with a high production cross section is $^{112}$Rh. The nucleus $^{112}$Rh contains two states both of which are populated in fission, the ground state with a half-life of 3.8 s and an isomeric state with a half-life of 6.8 s \cite{23}. Additionally the $^{112}$Rh mother nucleus, $^{112}$Ru, also produced in the fission reaction can feed into the ground state of $^{112}$Rh. The two states subsequently beta decay into excited states of $^{112}$Pd which de-excite emitting gamma rays. By analyzing the resulting gamma ray spectrum one can obtain an indication of the ion guide performance \cite{34}. As the ion guide efficiency is extracted from a measurement of the decay of $^{112}$Rh previous data from ``IGISOL 3'' exists in order to make a skimmer-SPIG comparison.

\begin{figure}[h]
\begin{center}
\includegraphics*[width=9cm]{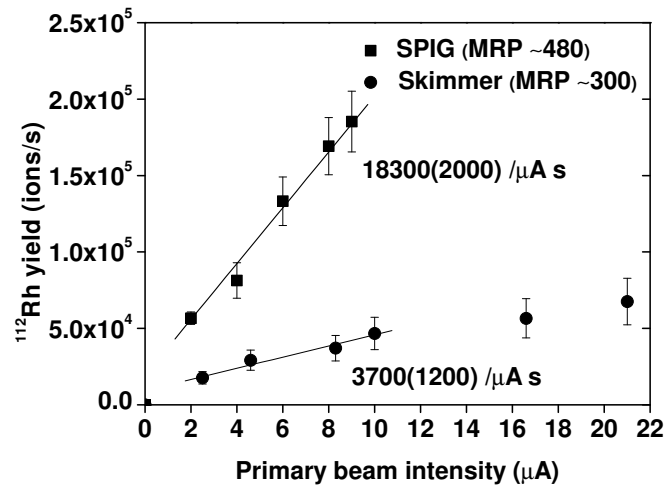}
\end{center}
\caption{$^{112}$Rh yields produced by 30 MeV proton-induced fission of $^{238}$U. The solid line represents the yield per $\mu$A of primary beam current estimated from a linear fit to the data.}
\label{fig:14}
\end{figure}

Figure 14 shows the result of the comparison as a function of primary beam intensity. In both cases, the helium pressure inside the ion guide was fixed at 200 mbar. The data for the SPIG are limited to 9 $\mu$A primary beam intensity due to the radiation safety limit of the IGISOL working area. In this work the mass resolving power of the SPIG system was approximately 1.6 times higher than that of the skimmer system. Variations in the mass resolving power have often been measured between different ion guides and depend on the operating conditions of the IGISOL front-end. A linear fit to the data in Fig. 14 shows that the SPIG provides on average 5 times higher yield of $^{112}$Rh as compared to the skimmer with same primary beam current. At the highest primary beam currents the skimmer data appears to deviate from the linear trend. This is an indication that space charge is starting to play a role (see discussion in section 5).

\subsection{Heavy-ion fusion-evaporation reaction}

In December 2005 the heavy-ion ion guide was installed at IGISOL for tests with the laser ion source \cite{24}, followed by an experiment at JYFLTRAP to measure masses of neutron-deficient nuclei close to the \textit{N=Z} line, important for the astrophysical rapid-proton capture (rp) process \cite{25}. During the laser ionization tests the skimmer and SPIG were installed during the same run and compared in the heavy-ion fusion-evaporation reaction of $^{nat}$Ni ($^{32}$S$^{7+}$,5p3n)$^{82}$Y. This reaction had previously been studied using the HIGISOL technique at IGISOL \cite{26} for decay spectroscopy measurements and therefore for a direct comparison with the earlier yields the previous reaction energy of 165 MeV was chosen for this work. The natural nickel target thickness was 3.7 mg/cm$^{2}$, consistent with that used previously. A setup on the central line of the IGISOL mass separator consisted of an MCP detector and a beta-gamma station with a plastic 3$\pi$ scintillator, a germanium detector and a tape drive for implantation of activities.

\begin{figure}[h]
\begin{center}
\includegraphics*[width=9cm]{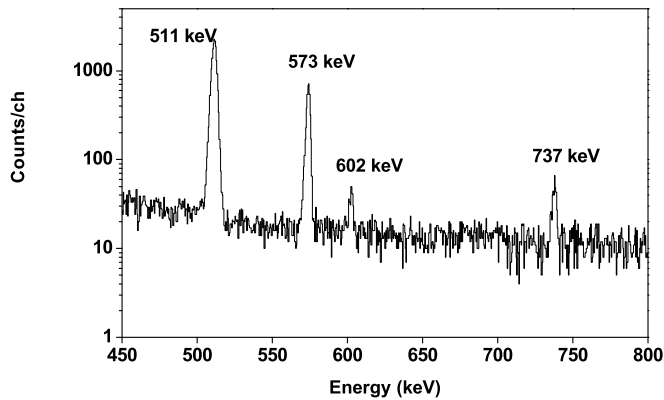}
\end{center}
\caption{A part of the beta-gated gamma spectrum from \cite{24}. Transitions related to the decay of $^{82}$Y are clearly visible - see text for details.}
\label{fig:15}
\end{figure}

With a half-life of $\sim$8.3 s, $^{82}$Y beta decays into excited states of $^{82}$Sr. A proposed decay scheme for $^{82}$Y can be found in \cite{26}. Part of the summed beta-gated gamma spectrum for \textit{A} = 82 radioactivity including both skimmer and SPIG data, taken from the work of \cite{24}, is shown in Fig. 15. The transitions of 573, 602 and 737 keV belong to the decay of $^{82}$Y. The beta-gated 573-keV gamma peak was used to extract a yield of 1.0$\pm$0.2 ions/s/pnA while using the skimmer. These nuclei were produced using a primary beam intensity of 28.6 pnA ($\sim$200 enA) and an ion guide helium pressure of 200 mbar. The effect of changing to the SPIG is shown in Fig. 16. It is immediately clear that in the situation when a HIGISOL reaction is used the SPIG does not improve the yield compared to the skimmer. The calculated yield of $^{82}$Y using the 573-keV peak is 1.0$\pm$0.3 ions/s/pnA with the SPIG.

\begin{figure}[h]
\begin{center}
\includegraphics*[width=9cm]{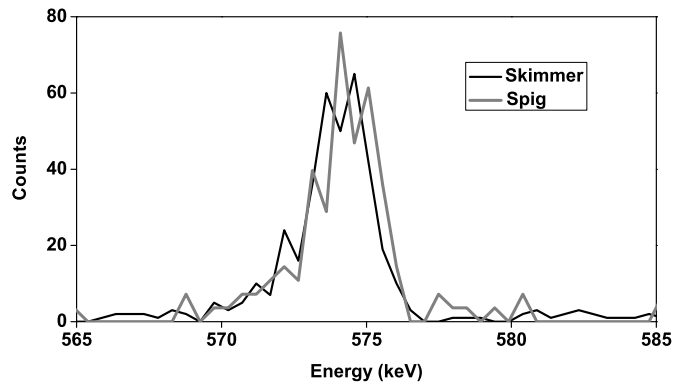}
\end{center}
\caption{Beta-gated 573-keV gamma-ray peak belonging to the decay of $^{82}$Y measured using the SPIG and the skimmer. Data for the skimmer was taken over a period of 70 minutes while that of the SPIG run was for 19 minutes. The SPIG data has been normalized to correspond to a 70 minute measurement.}
\label{fig:16}
\end{figure}

A comparison can be made with the yield presented in this work and that given in \cite{26}, in which an earlier version of the HIGISOL guide had been used. In the earlier work a maximum beam intensity of 125 pnA resulted in an $^{82}$Y yield of 260(70) ions s$^{-1}$, therefore $\sim$2 ions/s/pnA. Although this is a factor of two higher than the present work, these earlier yields had been measured after several days of beam time. It is well-known that the heavy-ion ion guide improves in performance over time. Additionally, in the recent work the background pressure in the IGISOL chamber was rather poor and therefore may have led to losses of $^{82}$Y$^{+}$ through molecular formation, as recently discussed in \cite{5}.

\subsection{Time of flight measurements and relationship to molecular formation}

Recent measurements showed that after exiting the ion guide, atomic yttrium ions can be quickly redistributed into a molecular form when the baseline pressure is at a level of between 5$\cdot$×10$^{-3}$ and 10$^{-2}$ mbar \cite{5}. As part of the on-going laser ion source programme, it is of interest therefore to determine the time of flight of ions through the SPIG and to see whether the timescale is feasible for molecular reactions to occur when the environment is poorly controlled. In order to study the time of flight the light-ion ion guide was installed with a natural nickel target of thickness 2.8 mg/cm$^{2}$. An 18 MeV proton beam was used to recoil target material into the helium gas which then acted as a source of ions in on-line conditions. The ions were mass separated and detected on a set of channel plates downstream from the focal plane of the separator. To study the time distribution of the ions, the ion signal from the channel plates was fed into a multi-channel analyzer (MCA) with a time resolution of 2.56 $\mu$s per bin, triggered by the JYFLTRAP labview control program \cite{27}. The program was used to control the timing of pulses sent to the repeller and end electrodes of the SPIG such that the dc potentials could be raised or lowered with respect to the ion guide potential. The delay times between the trigger signals from the labview program and the switching of the electrodes were less than 2 $\mu$s.

\begin{figure}[ht]
\begin{center}
\includegraphics*[width=9cm]{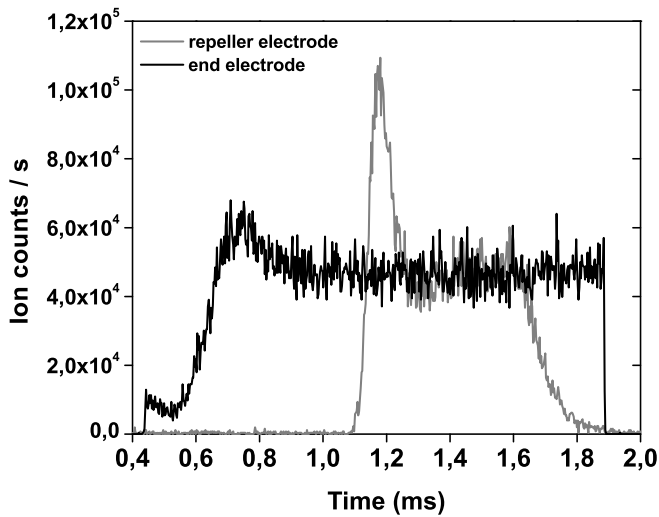}
\end{center}
\caption{Time-of-flight profiles of mass-separated nickel ions. The solid black line shows the time profile of ions after the end electrode potential is lowered with respect to the dc potentials on the SPIG rods after 400 $\mu$s, and is raised after 1.85 ms. The solid grey line shows the time profile of ions after the repeller electrode potential is lowered with respect to the ion guide after 1 ms and raised at 1.5 ms.}
\label{fig:17}
\end{figure}

A typical time distribution profile is illustrated in Fig. 17 taken with a helium gas pressure of 100 mbar within the ion guide. The solid black line shows the ion distribution after the end electrode potential is lowered with respect to the dc potentials on the SPIG rods, 400 $\mu$s after the start trigger of the scan. In this instance, the repeller electrode is at a nominal potential for ion transportation. There is a delay of $\sim$36 $\mu$s until the first ions are detected, the count rate then rises quickly and overshoots. These initial ions represent those that are extracted on-axis through the SPIG. The arrival of the majority of ions, those not extracted on the SPIG axis, occurs at a later time. The count rate peaks at $\sim$60000 ions s$^{-1}$ after 750 $\mu$s and then reduces to a steady-state rate of $\sim$50000 ions s$^{-1}$. This peak may well reflect the ions that have been accumulating within the SPIG while the end electrode potential was high with respect to that of the SPIG rods. The dc potential on the end electrode is raised after 1.85 ms and $\sim$36 $\mu$s later the ion signal drops rapidly within approximately 3 $\mu$s.

The corresponding ion distribution in time when the repeller electrode is pulsed is shown as the grey solid line in Fig. 17. In this instance the end electrode potential is set to allow ions to exit the SPIG. A delay of 100 $\mu$s exists between the repeller electrode switching time at 1 ms and the arrival of the first ions. Initially the transport efficiency through the SPIG is high, however as more charge enters the SPIG space charge effects occur. It is likely that the losses due to space charge determine the amplitude of the peak seen at $\sim$1.2 ms. For the fixed primary beam intensity used in this experiment the equilibrium count rate is determined by the SPIG transport efficiency. The repeller electrode potential is raised after 1.5 ms following which the count rate is maintained for a further 130 $\mu$s before reducing over a timescale far slower than the initial rise time. The difference in the rise time and fall time after the repeller potential is switched is attributed to the effects of space charge. In equilibrium the space charge acts to shield the full dc potentials of the SPIG rods, resulting in a slower flight time through the SPIG.

The ion time profiles of Fig. 17 can be used to estimate a flight time through the SPIG for the conditions used in this experiment. The data associated with the switching of the repeller potential are differentiated and a simple Gaussian is fitted to the resultant peak related to the arrival of the main bulk of the ions. The trigger time from the JYFLTRAP control program is then subtracted from the time obtained via the fitted peak and, lastly, a flight time through the mass separator is subtracted. This has been estimated to be $\sim$36 $\mu$s from the data obtained after switching the end electrode potential. The extracted flight time through the SPIG has been estimated to be $\sim$100 $\mu$s.

Additionally we note that the time profiles can be used to illustrate the sensitivity to changes in the environment, for example when introducing a leak into the baseline vacuum chamber as discussed in \cite{5}. However, the results obtained can be explained only qualitatively and therefore will not be discussed here. It is of interest to combine the extracted time of flight through the SPIG from this work with data published from the IGISOL laser ion source \cite{5}. In that work the ion guide was operated at a gas pressure of 150 mbar. In the system with no leak added to the IGISOL vacuum chamber the count rate of atomic yttrium \textit{Y} was measured to be $\sim$8200 ions s$^{-1}$. The only impurity seen was yttrium oxide at a level of $\sim$300 ions s$^{-1}$. From the time profiles it could be determined that the yttrium oxide was formed within the gas cell. When a leak was added to the vacuum chamber in \cite{5}, the atomic yttrium was redistributed into other molecular forms on fast timescales in the gas jet environment after the ion guide. The mass-separated yield of yttrium was reduced to $\sim$300 ions s$^{-1}$. We can use the simple decay law relationship $Y = Y_{0} \cdot exp(-k[M]t)$, where $Y_{0}$ represents the atomic yttrium entering the SPIG, \textit{k} is the chemical reaction rate coefficient (cm$^{3}$ s$^{-1}$) and [\textit{M}] the molecular concentration, to estimate a value for the molecular reaction time constant $\tau$ which is inversely proportional to the reaction rate \textit{k}[\textit{M}]. With an initial rate of 8200 ions s$^{-1}$, a final measured rate of 300 ions s$^{-1}$ and  a flight time \textit{t} through the SPIG of 100 $\mu$s the molecular reaction time constant $\tau$ is $\sim$30 $\mu$s. This is well within the flight time through the SPIG and so confirms the need for a clean environment (corresponding to a good baseline vacuum pressure) after the ion guide, in particular for very chemically reactive elements.

\subsection{SPIG transmission efficiency}

In January 2008 an experiment was performed to determine the transmission efficiency of the SPIG as a function of ion guide current. As a secondary interest, the output current of an ion guide as a function of the primary beam current was studied. In this test the light-ion ion guide was installed and a proton beam of 40 MeV impinged onto a 4.3 mg/cm$^{2}$ thick magnesium target. Due to the reaction kinematics the primary beam penetrating the stopping volume cannot be avoided. A single proton deposits $\sim$2.1 keV into 3 cm thickness of helium gas operating at 300 mbar. The energy deposited by the reaction products and target recoils can be ignored due to several orders of magnitude reduction compared to the number of protons passing through the ion guide. With an ion guide volume of 3 cm$^{3}$, the number of ion-electron pairs created per proton can be estimated to be $\sim$17. The total ionization-rate density \textit{Q} is proportional to the primary beam current.

\begin{figure}[h]
\begin{center}
\includegraphics*[width=9cm]{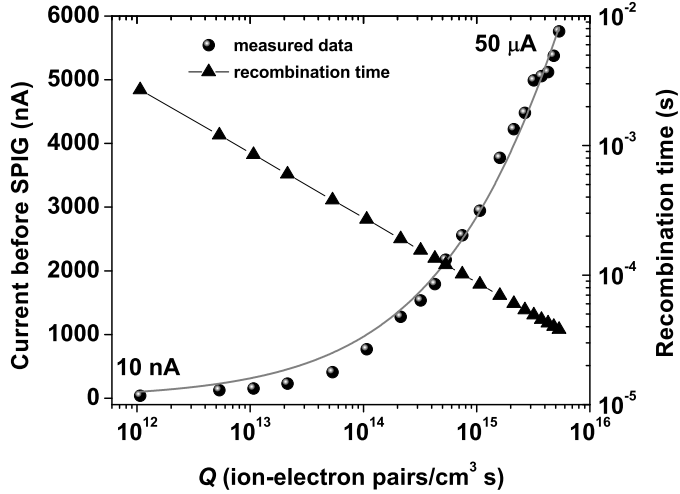}
\end{center}
\caption{The current measured before the SPIG as a function of ionization-rate density in the ion guide together with a fit of Eq. (2) which describes the time evolution of the charge density (solid spheres). The equivalent proton primary beam intensities are shown for the lowest and highest values of ionization-rate density. The associated recombination time $\tau$ is shown for each value of Q (solid triangles).}
\label{fig:18}
\end{figure}

Figure 18 illustrates the current obtained on a Faraday cup before the SPIG as a function of the ionization-rate density within the ion guide for primary beam current intensities ranging from 10 nA to 50 $\mu$A. The Faraday cup was biased to a voltage of -100 V at which a saturation of detected current was obtained. By plotting the data as a function of primary beam intensity on a linear scale the output current is seen to reveal a tendency towards a saturation value. The time evolution of the ion-electron density from the point of creation in the ion guide is given by n(t) \cite{28} where

\begin{equation}
n(t)=\frac{n_{0}}{1+\alpha\cdot n_{0}t}. 
\end{equation}

Here \textit{n}$_{0}$ is the initial density defined by $(Q/\alpha)^{1/2}$, \textit{t} represents an evolution time within the ion guide and $\alpha$ is the recombination rate coefficient of helium \cite{29}. The fit to the data in Fig. 18 is that of Eq. (2) normalized by a scaling factor. An evolution time of $\sim$10 $\mu$s was extracted from the chi-squared minimization routine. It appears that above an ionization-rate density of $2\cdot10^{14}$ pairs/cm$^{3}$s (corresponding to $\sim$2 $\mu$A primary beam intensity) the function fits the data well indicating that recombination becomes the driving mechanism of the trend towards saturation of the output current. Below $2\cdot10^{14}$ pairs/cm$^{3}$s the function over-estimates the experimental data. This can be understood with help from the second set of data in Fig. 18, indicated by the solid triangles. For every data point associated with a specific ionization-rate density \textit{Q}, a recombination time $\tau$ can be calculated using $1/(Q\cdot\alpha)^{1/2}$. The recombination timescales associated for the ionization-rate densities above $2\cdot10^{14}$ pairs/cm$^{3}$s are below 200 $\mu$s. The total evacuation time of the ion guide is a few ms which, to be in competition with the recombination time, corresponds to the lowest ionization-rate densities measured in this work. At low primary beam intensities the recombination time is no longer the dominant loss mechanism and the output current from the ion guide follows a simple linear increase with primary beam current. When the recombination timescale becomes the dominant ``loss factor'' against ion survival the output current follows a square root dependence of the ionization-rate density.

\begin{figure}[h]
\begin{center}
\includegraphics*[width=9cm]{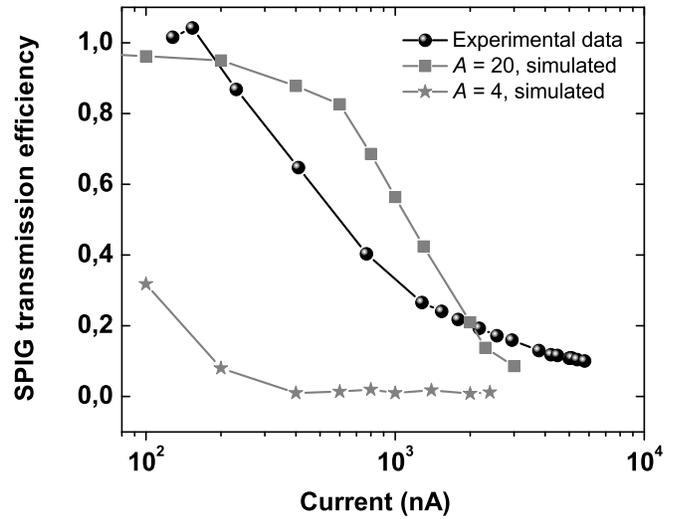}
\end{center}
\caption{The transmission efficiency of the SPIG as a function of ion guide beam current. The experimental data is shown compared with simulations of \textit{A} = 4 and \textit{A} = 20 test masses.}
\label{fig:19}
\end{figure}

Figure 19 illustrates the transmission efficiency of the SPIG as a function of beam current measured using the Faraday cup before the SPIG and a cup before the mass separator. Care should be taken when interpreting this data as the beam is accelerated to 30 kV before reaching the second Faraday cup. We assume that any secondary effects due to the beam impinging on the second cup are negligible. At currents of $\sim$0.1 $\mu$A the SPIG can transmit 100\% of the incoming charge density. However, as the current increases the transmission drops steadily before tending towards an equilibrium level after a few $\mu$A. A 50\% transmission efficiency level is reached after $\sim$0.6 $\mu$A.

The results of two test mass simulations are also shown, both performed at 300 mbar ion guide pressure, 250 $V_{pp}$ rf amplitude and realistic dc potentials on the electrodes. The first is \textit{A} = 20 whose trend closely follows that of the simulation performed at 100 mbar ion guide pressure in Fig. 11. In the region between 200 nA and 2 $\mu$A the simulation efficiency is overestimated compared with experiment and then underestimated at currents above $\sim$2 $\mu$A. This mass reaches a 50\% transmission efficiency level at $\sim$1.1 $\mu$A. It is known that helium is not easily transported through the SPIG and it has never been seen as a dominant peak after mass separation. This is reflected in the second simulation, that of \textit{A} = 4, whose transport efficiency drops rapidly to $\sim$1\% at less than 0.5 $\mu$A beam current. The experimental data reflects the overall trend of transmission through the SPIG as a function of current, yet a more informative comparison to the simulations will be made in the future using mass-separated beams. From the present work we can conclude that the SPIG can efficiently transport $\sim$10$^{12}$ ions s$^{-1}$ with minimal losses however above this rate space charge effects occur and the transmission efficiency starts to decrease.

\section{Discussion}

Off-line, the measured efficiencies of $^{219}$Rn$^{+}$ using the skimmer and SPIG devices as a function of ion guide pressure are rather similar (Fig. 12). This does not reflect the data seen in Fig. 14(b) of \cite{11} in which only three data points for the SPIG system were obtained, and appear to show that at low pressures ($\sim$80 mbar) the transmission efficiency of the SPIG is significantly higher than the skimmer-ring system ($\sim$4$\times$ improvement). At higher pressures close to 400 mbar the efficiencies become comparable. In the off-line transmission tests at Leuven, a 10\% increase in transmission efficiency was reported compared to the skimmer with the same power deposition in the nickel filament \cite{10}. Although the simulations in this paper have not been performed for masses heavier than \textit{A} = 106, the trend in Fig. 11 supports the idea that the SPIG transmission should be slightly higher than that of the skimmer at pressures close to 100 mbar for small ion guide currents ($<$10 pnA). At higher pressures the skimmer efficiency may start to be reduced due to collisional scattering between the extracted ions and the buffer gas atoms.

A summary of the on-line yields obtained in this work is presented in Table 2. The increase of a factor of five in the fission yield of $^{112}$Rh$^{+}$ with the SPIG compared to the skimmer system is similar to that reported in \cite{11}. In that work the production rate of $^{112}$Rh$^{+}$ was 14700 ions s$^{-1}$ with a 1 $\mu$A proton beam and, with 4.5 $\mu$A, 7400 ions/$\mu$C. A drop in efficiency at higher primary beam intensity was seen to be of the same order for both the skimmer and SPIG, and it was concluded that this reflected loss processes happening within the gas cell. Similarly, a SPIG was tested in an experiment using proton-induced fission of uranium to produce $^{69}$Ni at the Leuven isotope separator \cite{10}. The intensity of the mass-separated $^{69}$Ni beam was approximately 5 times higher compared to that with the skimmer, at the same ion guide pressure (500 mbar). However, with other improvements to the IGLIS and the separator, the authors could not conclude that this gain in efficiency was due to the SPIG alone.

An increase in the yields of isotopes produced in light-ion induced fusion-evaporation reactions, using the SPIG, has been seen not only in the two cases listed in Table 2, but across a variety of elements and mass regions \cite{21,30}. Interestingly, in heavy-ion induced fusion-evaporation reactions there is no increase in yield (Table 2). This has also been seen and reported in \cite{11}, for the reaction $^{96}$Mo($^{36}$Ar,3p2n)$^{125}$La. The yield of $^{125}$La was reported to only have a slight improvement with the use of the SPIG (a factor of 1.3). As the primary beam is prevented from entering the stopping chamber in a heavy-ion reaction the current extracted from the ion guide is rather low, typically a few tens of nA. This can be compared to fission in which a proton bombarding intensity of several $\mu$A leads to an extracted current of a few hundred nA and, in light-ion induced fusion-evaporation reactions, a few $\mu$A of primary beam leads to almost 1 $\mu$A of detected current. The lack of clear improvement in a heavy-ion reaction is in fact suggested by the simulations of Fig. 11 in this work. As indicated, at low ion guide currents the ratio of the transmission efficiency of the SPIG compared to the skimmer reduces for heavier masses (in Fig. 11 the ratio of improvement is a factor of 2 for \textit{A} = 20 yet only $\sim$1.2 for \textit{A} = 106). On the other hand, Fig. 5 illustrates the sensitivity of the skimmer efficiency to the ion guide pressure. The pressure in the region between the skimmer and extractor electrode would need to be accurately measured before drawing firm conclusions.

\begin{table}[h]
\caption{A summary of the on-line yields obtained in this work with the SPIG system compared to the skimmer system.}
\label{tab:2}
\begin{tabular}{|p{1.1cm}|p{2.8cm}|p{1.3cm}|p{1.3cm}|p{1.3cm}|p{0.8cm}|}
\hline
Element & Reaction & Ion guide & SPIG [ions/$\mu$C] & Skimmer [ions/$\mu$C] & SPIG gain \\
\hline
$^{40}$Sc & $^{40}$Ca(p,n)$^{40}$Sc & Light ion & 109 & 14 & 7.8 \\
$^{58}$Cu & $^{58}$Ni(p,n)$^{58}$Cu & Light ion & 3200 & 267 & 12.0\footnotemark[1]   \\
$^{82}$Y & $^{nat}$Ni($^{32}$S,5p3n)$^{82}$Y & HIGISOL & 1000(300) & 1000(200) & 1 \\
$^{112}$Rh & $^{238}$U(p,fission)$^{112}$Rh & Fission & 18250 & 3656 & 5.0 \\
\hline
\end{tabular}
\vspace*{2pt}
\end{table}

\footnotetext[1]{The gain in efficiency with the SPIG is given with respect to data taken with ``IGISOL-2'' [21]. Typically a factor of three improvement in the yields was reported in \cite{22} after the upgrade to ``IGISOL-3''. Therefore the real improvement using the SPIG compared to the skimmer-ring system is a factor of 4.}

It is of interest to compare the space charge simulation results of Fig. 11 with that of the experimental data obtained in section 4. The $^{40}$Sc$^{+}$ yields produced by the $^{40}$Ca(p,n)$^{40}$Sc reaction are shown as a function of primary beam intensity in Fig. 13. In this reaction a proton of energy 35 MeV deposits 1.2 keV into 3 cm thickness helium gas at an ion guide pressure of 150 mbar. With an ion guide volume of 3 cm$^{3}$, the number of ion-electron pairs created per proton is $\sim$10. There are two methods we have used in order to estimate the ion guide current which can then be compared directly with the simulation results of Fig. 11. The first method assumes an equilibrated charge density inside the ion guide volume. For each value of primary beam intensity used in Fig. 13 an associated ionization-rate density \textit{Q} (pairs/cm$^{3}$ s) can be calculated. The corresponding equilibrium density \textit{n}$_{0}$ (pairs/cm$^{3}$) is then multiplied by the exit nozzle area (1.2 mm diameter) and the gas velocity at the nozzle (1000 ms$^{-1}$) to provide a simple estimate of the current exiting the ion guide. In order to check the validity of this simple model we have estimated the current for the data shown in Fig. 18 and compared it to the measured Faraday cup current. The results are shown in Fig. 20 along with a linear fit to the data simply used to guide the eye. The fit has not been weighted with the error bars which are taken to be conservative 10\% values of the measured data.

The conclusion that can be drawn from the model is that it overestimates the measured data by a factor of $\sim$6. In the same figure we also show the ratio of the measured current to estimated current extracted from the ion guide. Ignoring the absolute value of the ratio it is interesting to note that the model predicts the equilibrium of the production of ion-electron pairs with the subsequent recombination above a \textit{Q} value of $\sim2\cdot10^{14}$ pairs/cm$^{3}$s. Below this ionization-rate density the model is no longer sufficient because the recombination time becomes equivalent to or even longer than the evacuation time of the ion guide (see the discussion in section 4.6).
	
\begin{figure}[h]
\begin{center}
\includegraphics*[width=9cm]{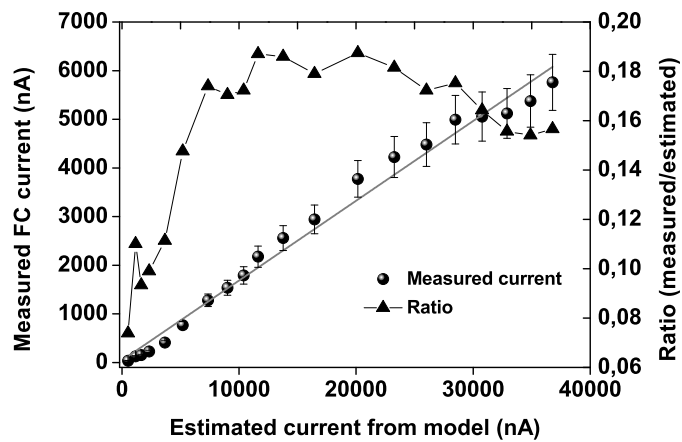}
\end{center}
\caption{The current measured before the SPIG compared to the estimated ion guide current with a linear fit to guide the eye (solid spheres). The ratio of the measured current to estimated current is shown as solid triangles.}
\label{fig:20}
\end{figure} 

Fortunately, despite the failure of the model to accurately predict the measured ion guide current, the data shown in Fig. 18 can be used as a calibration curve to convert the associated ionization-rate densities of the light-ion reaction (Fig. 13) into an accurate value of the ion guide current. This cannot be done for other experimental data in this work, for example that of the proton-induced fission of $^{238}$U, because an equivalent study of the ion guide current for that particular gas cell has not been performed. A summary of the data from Fig. 13 is plotted in Fig. 21 as a function of ion guide current, along with the overall experimental SPIG transmission reproduced from Fig. 19 and a simulation of the transmission efficiency for \textit{A} = 20 for both the skimmer and SPIG. The relative efficiency of $^{40}$Sc$^{+}$ is calculated from the measured yield (ions/s) per microampere of primary beam current, normalized to the maximum value for both the skimmer and SPIG data respectively. All other data are treated in a similar way in order to show the relative efficiency.
It is striking to see how the combination of all the data sets, measured via light-ion induced fusion-evaporation yield studies and simulations, show a relatively similar overall trend in decreasing relative (transmission) efficiency as a function of ion guide current, with the 50\% efficiency level clustered around 1 $\mu$A ion guide current. It is not surprising to see differences between the simulations and experiment (for example how sharply or slowly the transmission reduces) due to unknown parameters such as a direct measurement of the pressure after the ion guide. From the measured data a transmission efficiency of 50\% is reached at 1 $\mu$A for the SPIG and, for the equivalent beam current, the skimmer only has an efficiency of 20\%. In general the agreement between experiment and simulation is very encouraging. For the first time we have been able to simulate the transmission efficiency as a function of current for the SPIG system and to directly compare this with experimental data.

\begin{figure}[h]
\begin{center}
\includegraphics*[width=9cm]{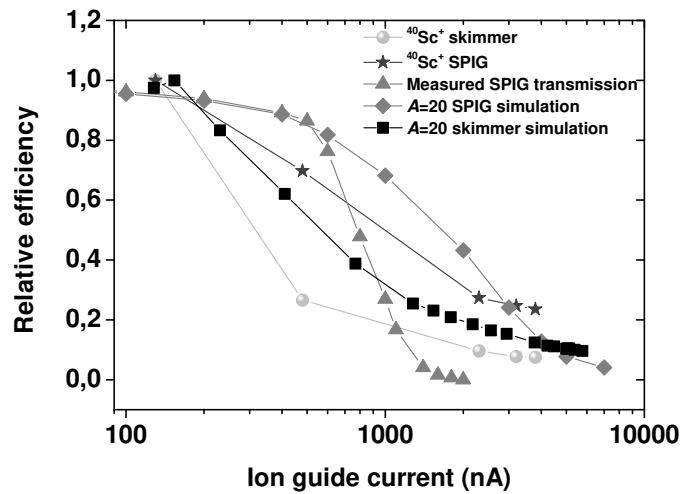}
\end{center}
\caption{Relative efficiency as a function of ion guide current. Solid squares show the measured SPIG transmission data of Fig. 19, solid spheres the relative efficiency of $^{40}$Sc$^{+}$ using the skimmer electrode, stars illustrate the relative efficiency of $^{40}$Sc$^{+}$ using the SPIG, solid triangles show the simulation of the SPIG transmission efficiency for \textit{A} = 20 reproduced from Fig. 11 and solid diamonds the skimmer simulation efficiency for \textit{A} = 20.}
\label{fig:21}
\end{figure} 

Finally we turn our attention to the discussion of the mass resolving power. This has been briefly discussed in section 3.4 in which, for negligible ion guide current, the SPIG was predicted to have a mass resolving power nine times higher than of the skimmer. In reality the mass resolving power has often varied depending on the type of ion guide used and the experimental conditions. The improvement of beam quality with the SPIG has been rarely measured to be more than a factor of three higher than compared to the skimmer. This is reflected in the simulated data shown in Fig. 22. The ratio of the standard deviation in the kinetic energy ($\Delta$KE) multiplied by the root mean square value of the radial axis (X$_{rms}$) for the skimmer compared with the SPIG (estimated at the position of the extractor electrode) is plotted as a function of ion guide current. At low values of ion guide current the ratio is $\sim$10. This reflects a combination of higher energy spread and larger emittance of the skimmer electrode, thus a factor of 10 improvement of the mass resolving power of the SPIG. For all values of simulated current the SPIG has a higher mass resolving power, yet the improvement factor reduces as the current increases. In the majority of the experiments at IGISOL (light-ion induced reactions and proton-induced fission) the measured currents before the mass separator are typically several hundreds of nA.

From the conversion of ionization-rate density to ion guide current (Fig. 18) a comparison of the simulated mass resolving power can be made with the measurements for the light-ion induced reaction of Fig. 13. The mass resolving power was measured in the focal plane of the mass separator at mass \textit{A} = 40 (argon) using a primary beam current of 1 $\mu$A. The SPIG had an improvement factor of 2.6 compared to the skimmer. A primary proton beam current of 1 $\mu$A results in an ion guide current of $\sim$770 nA which can be used to extract a simulated SPIG improvement factor of $\sim$3.7 from Fig. 22. Even though the pressure within the SPIG is unknown and the simulated mass resolving power is sensitive to this parameter, the comparison between simulation and experiment is once again very encouraging.

\begin{figure}[h]
\begin{center}
\includegraphics*[width=9cm]{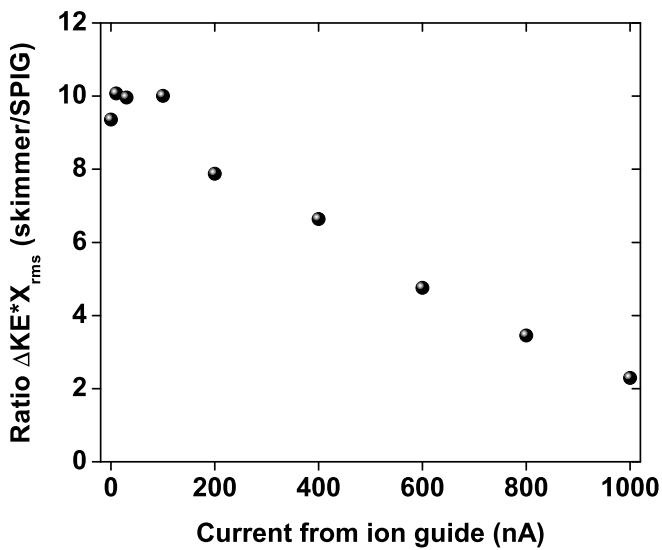}
\end{center}
\caption{The ratio of the standard deviation in the kinetic energy ($\Delta$KE) multiplied by the root mean square value of the radial axis (X$_{rms}$) for the skimmer compared with the SPIG, as a function of ion guide current. This ratio is inversely proportional to the mass resolving power.}
\label{fig:22}
\end{figure} 

\section{Conclusions}

In summary we have built an rf sextupole ion beam guide (SPIG) for use at IGISOL, motivated by the development of the laser ion source and the goal to achieve extremely high selectivity in the production of exotic radioactive ion beams. In the present work we have concentrated on the use of the SPIG as an ion transport device. Simulations have been performed to understand the differences between the transportation of ions through the skimmer and the SPIG system, with and without space charge effects. One of the more important parameters in the model that has not been measured directly is that of the pressure between the ion guide, the skimmer/SPIG device and the extraction electrode. In principle measurements of the pressure distributions could be made in the future using thin static and stagnation pressure probes as previously discussed by Iivonen \textit{et al} \cite{32}. 

The effects of space charge have been estimated in a rather simple model. The radius of the charge cylinder used is independent of the beam current, which is unrealistic and directly leads to an overestimation of the charge density with higher current beams. Correspondingly this results in an underestimation of the transmission efficiency. Most recently more advanced simulations have been performed which simulate the trajectories of 500 test ions simultaneously. Although this needs more computing time it gives the possibility to recalculate the beam current in a realistic manner each time an ion is lost. The saturation of the output current is clearly seen and is shown to scale with a $V_{rf}^{2}$ dependence. This agrees with more detailed simulations performed in \cite{33}.
	
For a direct comparison of the simulations with different ion guides and production reactions it will be necessary to directly measure the output current from each ion guide as a function of primary beam intensity. However, the good agreement with data from the light-ion induced fusion reaction presented in this work suggests we have a good understanding of the transport mechanisms. In future measurements of ion guide efficiencies it will be important to deconvolute the extraction efficiency from the gas cell from that of the transmission through the SPIG. The saturation of the ion guide current is of primary interest as it indicates possible limitations in the ion guide technique. Experiments are currently underway to study this effect further both at room temperature and cryogenic temperature conditions.

\section*{Acknowledgments}
The authors wish to thank Dr. Andrey Popov for enlightening discussions. This work has been supported by the LASER Joint Research Activity Project under the EU 6$^{th}$ Framework program ``Integrating Infrastructure Initiative-Transnational Access'', Contract number: 506065 (EURONS) and by the Academy of Finland under the Finnish Centre of Excellence Program 2006-2011 (Nuclear and Accelerator Based Physics Program at JYFL).

\end{document}